\documentclass[aps,prl,twocolumn,amsmath,amssymb,amsfonts,nofootinbib,long]{revtex4}
\usepackage{epsfig,latexsym,bm}

\newcommand\eV{\mbox{eV}}

\newcommand\MeV{\mbox{MeV}}
\newcommand\GeV{\mbox{GeV}}

\newcommand\AU{\mbox{AU}}
\newcommand\pc{\mbox{pc}}
\newcommand\kpc{\mbox{kpc}}

\newcommand\G{\mbox{G}}
\newcommand\A{\mathbf{A}}
\newcommand\B{\mathbf{B}}
\newcommand\E{\mathbf{E}}
\newcommand\x{\mathbf{x}}
\newcommand\y{\mathbf{y}}
\newcommand\z{\mathbf{z}}
\newcommand\kk{\mathbf{k}}
\newcommand\p{\mathbf{p}}
\newcommand\0{\mathbf{0}}
\newcommand\mPl{m_{\rm Pl}}
\newcommand\dd{\partial}
\newcommand\rot{\nabla \times}
\newcommand\e{\mathbf{e}}

\begin{document}

\title{Lorentz Symmetry Violation and Galactic Magnetism}

\author{L. Campanelli$^{1,2}$}
\email{leonardo.campanelli@ba.infn.it}
\author{P. Cea$^{1,2}$}
\email{paolo.cea@ba.infn.it}
\author{G.L. Fogli$^{1,2}$}
\email{gianluigi.fogli@ba.infn.it}

\affiliation{$^1$Dipartimento di Fisica, Universit\`{a} di Bari, I-70126 Bari, Italy}
\affiliation{$^2$INFN - Sezione di Bari, I-70126 Bari, Italy}

\date{December, 2008}


\begin{abstract}
We analyze the generation of primordial magnetic fields during de
Sitter inflation in a Lorentz-violating theory of Electrodynamics
containing a Chern-Simons term which couples the photon to an
external four-vector. We find that, for appropriate magnitude of
the four-vector, the generated field is maximally helical and,
through an inverse cascade caused by turbulence of primeval
plasma, reaches at the time of protogalactic collapse an intensity
and correlation length such as to directly explain galactic
magnetism.
\end{abstract}


\maketitle


Up to today, no compelling evidence for violation of Lorentz
symmetry exists (for a recent review see~\cite{Kostelecky}). The
main motivation to investigate possible consequences of Lorentz
violation is that in unified theories at the Planck scale, such as
quantum gravity, it is allowed or even unavoidable. Hence, any
direct or indirect evidence of Lorentz violation would give
information about very-high-energy physics or, in a cosmological
context, about physics of the very early universe.

In this paper, we indeed claim that the existence of galactic
magnetic fields could be an indirect evidence of Lorentz symmetry
violation. Magnetic fields are observed in any types of galaxies
(for reviews see~\cite{Widrow}). They are correlated on scales of
order of galactic dimensions ($\simeq 10 \kpc$) and have an
intensity of about $10^{-6} \G$.

In the existing literature, it is assumed that a primordial
magnetic field $B_\xi$ (coherent within domains of physical size
$\xi_{\rm phys}$) remains frozen into the plasma during the {\it
collapse} of a protogalaxy owing to the high conductivity of the
protogalactic gas~\cite{Widrow}. Accordingly, since the physical
size $L_{\rm phys} \simeq 10\kpc$ and mass density $\rho_{\rm pg}
\simeq 10^{-23} \mbox{g} \, \mbox{cm}^{-3}$~\cite{note1} of a
protogalaxy remain approximately constant during the collapse,
$B_\xi$ is amplified by a factor $[\rho_{\rm pg}/\rho(t)]^{2/3}$
while $\xi_{\rm phys}$ is reduced by $[\rho_{\rm
pg}/\rho(t)]^{1/3}$ with respect to a field which is not trapped
in a protogalaxy. Here, $\rho(t) \propto 1/\,t^2$ is the
(time-dependent) total energy density of the universe. Taking into
account that $\rho(t_0) \simeq 10^{-29} \mbox{g} \,
\mbox{cm}^{-3}$~\cite{Kolb}, where $t_0 \simeq 13.7 \, \mbox{Gyr}$
is the actual age of the universe~\cite{WMAP}, one find the
``canonical'' result that the overall amplification of $B_\xi$ and
diluition of $\xi_{\rm phys}$ are respectively $10^4$ and
$10^2$~\cite{Widrow}.

The above analysis, however, is not completely correct. In fact,
being the collapse of the protogalactic gas a turbulent
phenomenon~\cite{Kulsrud}, we expect that the magnetic field will
be processed by magnetohydrodynamic effects. Defining as usual the
kinetic and magnetic Reynolds numbers at the length scale
$\xi_{\rm phys}$ as $\mbox{Re} = v \xi_{\rm phys}/\nu$ and
$\mbox{Re}_B = v \xi_{\rm phys} \sigma_c$, where $v$ is the
typical velocity of the fluid motion, $\nu$ the kinematic
viscosity, and $\sigma_c$ the conductivity, magnetohydrodynamic
turbulence occurs when $\mbox{Re}$ and $\mbox{Re}_B$ are greater
than unity~\cite{Biskamp}. Applying the results of~\cite{Kulsrud}
we have that, in a typical protogalaxy, $v$ is given by the
``virial thermal speed'', $v_{\rm th} \simeq 10^{-3}$~\cite{R1},
the kinematic viscosity is dominated by ions, $\nu \simeq \nu_{\rm
i}
\simeq 10^{-4} \pc$, 
while the conductivity is the inverse of the
Spitzer resistivity, $\eta_S \simeq 10^{-25} \pc$. 
Consequently, we have $\mbox{Re} \simeq \xi_{\rm phys}/0.1\pc$ and
$\mbox{Re}_B \simeq 10^{21} \mbox{Re}$, 
which shows that magnetic domains of physical size greater than
$0.1\pc$ evolve turbulently during protogalactic collapse.

In the framework of freely-decaying magnetohydrodynamic
turbulence, it is a well-known result that the evolution of a
maximally helical field is characterized by the quasi-conservation
of magnetic helicity which, in turn, triggers and sustains a
transfer of magnetic energy from small to large scales, a
mechanism known as {\it inverse cascade}~\cite{Biskamp} (for a
definition of magnetic helicity and maximally helical fields, see
below). As a result, the intensity of the magnetic field inside a
domain is not longer constant but decays as $B_\xi \propto
\xi_{\rm phys}^{-1/2}$~\cite{Son}, while the physical correlation
length grows as $\xi_{\rm phys}(t) \simeq \xi_{\rm phys}(t_{\rm
pg}) [t/\tau_{\rm eddy}(t_{\rm pg})]^{2/3}$~\cite{Son} for $t \gg
\tau_{\rm eddy}(t_{\rm pg})$, where $t_{\rm pg}$ is the time when
protogalactic collapse begins (corresponding to a redshift of
order $z_{\rm pg} \sim 10^2$~\cite{Widrow,R2}) and $\tau_{\rm
eddy}(t) = \xi_{\rm phys}/v$ is the so-called eddy turnover time.
Using this latter evolution law, we find that after a time $t_*
\simeq [0.1 \pc/\xi_{\rm phys}(t_{\rm pg})]^{1/2} t_0$ the
correlation length reaches the maximum possible value
corresponding to the protogalaxy dimension $L_{\rm phys}$. This
means that for $t \gtrsim t_*$ the magnetic field remains
effectively frozen in the protogalaxy, $B_\xi(t) = \mbox{const}$
and $\xi_{\rm phys}(t) = \mbox{const}$. The foregoing analysis, if
on the one hand is oversimplified, on the other hand indicates
that if the magnetic field at the time of protogalactic collapse
is $i$) turbulent [i.e. $\xi_{\rm phys}(t_{\rm pg}) \gtrsim
0.1\pc$], $ii$) maximally helical, and $iii$) has an intensity
equal to $B_\xi(t_{\rm pg}) \simeq 10^{-6} [10 \kpc/\xi_{\rm
phys}(t_{\rm pg})]^{1/2} \G$ [so that $B_\xi(t_*) = 10^{-6} \G$],
it explains galactic magnetism.

The possibility that Lorentz violation could give rise to
large-scale magnetic fields was investigated for the first time by
Bertolami and Mota~\cite{Bertolami} in the context of string
theories. Developing the original idea by Kosteleck\'{y}, Potting
and Samuel~\cite{Kostelecky-p}, they showed that the appearance of
an effective photon mass owing to spontaneous breaking of Lorentz
invariance allows the generation of astrophysically interesting
magnetic fields within inflationary scenarios. To our knowledge,
the connection between Lorentz symmetry violation and cosmic
magnetic fields has been studied only in three other papers,
within the framework of either noncommutative
spacetimes~\cite{Mazumdar} or quantum theory with noncommutative
fields~\cite{Gamboa}.


In this paper, we consider a modification of standard
electromagnetism by adding to the Maxwell Lagrangian a
Chern-Simons term. Although this is not the only admissible
Lorentz-violating term~\cite{Kostelecky1}, it is the archetypal
one. First studied by Carroll, Field, and Jackiw~\cite{Carroll} in
1990, its cosmological and astrophysical implications are sill
investigated nowadays (see, e.g.,~\cite{WMAP}). In curved
spacetimes, the Maxwell-Chern-Simons action
reads~\cite{Kostelecky1}:
\begin{equation}
\label{Action} S_{\rm MCS} = \int \!\! d^4x \, e \! \left( -
\mbox{$\frac14$} \, F_{\mu \nu} F^{\mu \nu} - \mbox{$\frac12$} \,
p_\kappa e_{\mu}^{\;\kappa} A_{\nu} \widetilde{F}^{\mu \nu}
\right) \!,
\end{equation}
where $F_{\mu \nu} = \partial_{\mu} A_{\nu} - \partial_{\nu}
A_{\mu}$ is the electromagnetic field strength tensor,
$\widetilde{F}^{\mu \nu} = (1/2e) \, \epsilon^{\mu \nu \rho
\sigma} F_{\rho \sigma}$, $e$ the determinant of the vierbein
$e_{\mu}^{\;\kappa}$, and $\epsilon^{\mu \nu \rho \sigma}$ the
Levi-Civita tensor density. 
The external four-vector $p^\mu$ pick out a preferred direction in
spacetime and then breaks (particle) Lorentz
invariance~\cite{Carroll,Colladay,note added}. Moreover, the
Chern-Simons term violates CPT symmetry and, in general, conformal
invariance (which is an essential requirement for the generation
of seed magnetic fields in {\it spatially-flat} models of the
universe~\cite{Barrow}). However, if one wants to retain gauge
invariance, a supplementary condition on $p_\mu$ has to be
imposed~\cite{Carroll}: $D_\mu p_\nu - D_\nu p_\mu = \dd_\mu p_\nu
- \dd_\nu p_\mu = 0$, where $D_\mu$ is the spacetime covariant
derivative. In particular, the above condition is satisfied if
$p^\mu$ is a constant, or if $p^i = 0$ and $p^0$ is an arbitrary
function of time (Latin indexes run from $1$ to $3$, while Greek
ones from $0$ to $3$).

The equations of motion follow from action~(\ref{Action}),
$D_{\mu} F^{\mu \nu} = p_\mu \widetilde{F}^{\mu \nu}$, while the
Bianchi identities are $D_{\mu} \widetilde{F}^{\mu \nu} = 0$. We
assume that during inflation the universe is described by a de
Sitter metric $ds^2 = a^2(d\eta^2 - d \x^2)$, where $a(\eta)$ is
the expansion parameter, $\eta = -1/(aH)$ the conformal time, and
$H$ the Hubble parameter.
Let us introduce in the usual way the electric and magnetic fields
as $F_{0i} = -a^2 E_i$ and $F_{ij} = \epsilon_{ijk} a^2 B_k$. In
vector notation, the equations of motion read
\begin{equation}
\label{Maxwell1} \dd_\eta (a^2 \E) - \rot a^2 \B = m a^2 \B - \p
\times a^2 \E
\end{equation}
and $\nabla \cdot \E = -\p \cdot \B$, while the Bianchi identities
become $\partial_\eta (a^2 \B) + \rot a^2 \E = 0$ and $\nabla
\cdot \B = 0$, where $p^\mu = (m,\p)$.

From Eq.~(\ref{Maxwell1}), we readily get that the effect of $\p$
is just to cause a precession of the electric field vector.
Therefore, if the external four-vector takes the form $p^\mu =
(0,\p)$, the Chern-Simons term does not break conformal invariance
of electromagnetism so that, as shown long time ago by Turner and
Widrow~\cite{Turner}, no astrophysically interesting magnetic
fields can be produced during inflation. For this reason, we will
restrict our analysis to the case $p^\mu = (m,\0)$, with $m$ a
positive mass parameter which, eventually, depends on the
time. 
In this case, the external four-vector does not introduce any
preferred direction in space (but just a preferred direction in
time) and inflation-produced magnetic fields will appear
homogeneous and isotropic. This means that the two-point
correlation tensor of the magnetic field, $\langle B_i(\x) B_j(\y)
\rangle$, where $\langle ... \rangle$ denotes ensemble average, is
a function of $|\x-\y|$ only and moreover transforms as an $SO(3)$
tensor. In terms of the Fourier amplitudes of the magnetic field,
$\B_\kk = \int \! d^3 x \, e^{-i \kk \cdot \x} \, {\textbf
B}(\x)$, where $\kk$ is the comoving wavenumber, these conditions
translate into~\cite{Monin}:
\begin{equation}
\label{CorrelatorB} \langle (\B_\kk)_i (\B_{\kk'})_j \rangle =
\delta (\kk + \kk') \!\! \left[ \mathrm{P}_{\!ij}^{(\mathrm{S})}
\frac{\pi^2\mathcal{P}_k}{k^3} - \mathrm{P}_{\!ij}^{(\mathrm{A})}
\frac{\pi^2\mathcal{H}_k}{k^2} \right] \!\! ,
\end{equation}
where $\mathrm{P}_{\!ij}^{(\mathrm{S})} = (2\pi)^3(\delta_{ij} -
\hat{k}_i \hat{k}_j)$, $\mathrm{P}_{\!ij}^{(\mathrm{A})} =
i(2\pi)^3 \varepsilon_{ijl} \hat{k}_l$, $\delta_{ij}$ is the
Kronecker delta, $\hat{k}_i = k_i/k$, $k = |{\textbf k}|$, and
$\varepsilon_{ijk}$ is the totally antisymmetric tensor in tree
dimensions. The functions $\mathcal{P}_k$ and $\mathcal{H}_k$, the
so-called magnetic field and magnetic helicity power spectra,
denote the symmetric and antisymmetric parts of the correlator.

Let us define the average magnetic field on a comoving scale
$\lambda$ and the magnetic helicity density in a volume $V =
\lambda^3$ respectively as~\cite{Campanelli} $B_\lambda^2 =
\langle \, | \int \! d^3y \, W_\lambda(\x-\y) \, \B(\y)|^2
\rangle$ and $H_\lambda = \langle \, | \int \! d^3y \! \int \!
d^3z \, W_\lambda(\x-\y) W_\lambda(\x-\z) \, \A(\y) \cdot \B(\z) |
\, \rangle$, where $W_\lambda(\x) = (2\pi \lambda^2)^{-3/2}
e^{-|\x|^2/(2\lambda^2)}$ is a Gaussian window function. Then,
taking into account Eq.~(\ref{CorrelatorB}), we get
\begin{equation}
\label{Blambda1-H2} B_\lambda^2 = \int_{0}^{\infty}\! \frac{dk}{k}
\, W_\lambda^2(k) \, \mathcal{P}_k, \;\; H_\lambda =
\int_{0}^{\infty} \! \frac{dk}{k} \, W_\lambda^2(k) \,
\mathcal{H}_k,
\end{equation}
with $W_\lambda(k) = e^{-\lambda^2 k^2/2}$ the Fourier transform
of $W_\lambda(\x)$.
%
It is useful to introduce the so-called orthonormal helicity basis
$\{\e^+_\kk, \e^-_\kk, \e^3_\kk\}$, where $\e^{\pm}_\kk = -i
(\e^1_\kk \pm \, i \e^2_\kk)/\sqrt{2}$, $\e^3_\kk =
\widehat{\kk}$, and $\{\e^1_\kk, \e^2_\kk, \e^3_\kk\}$ form a
right-handed orthonormal basis. The Fourier transform of the
magnetic field can be then decomposed as $\B_\kk = B_\kk^+
\e_\kk^+ + B_\kk^- \e_\kk^-$, $B_\kk^\pm$ being the positive and
negative helicity components of $\B_\kk$, so the magnetic and
helicity power spectra become~\cite{Caprini} $\mathcal{P}_k =
(k^3/2\pi^2) \left(|B_\kk^+|^2 + |B_\kk^-|^2\right)$ and
$\mathcal{H}_k = (k^2/2\pi^2) \left(|B_\kk^+|^2 -
|B_\kk^-|^2\right)$.
%
%
%
%
%
A magnetic field is {\it maximally helical} at the time $\eta'$ if
either $B_\kk^+(\eta')$ or $B_\kk^-(\eta')$ is zero.

Taking the curl of Eq.~(\ref{Maxwell1}), using the Bianchi
identities, and going into Fourier space, we get
\begin{equation}
\label{Autoinduction} \dd^2_\eta (a^2 B_\kk^\pm) + k(k \pm m) \,
a^2 B_\kk^\pm  = 0.
\end{equation}
Before proceeding farther, we note that Eq.~(\ref{Autoinduction})
is valid also during radiation and matter dominated eras. Positing
in it the phase-exponential ansatz, $a^2 B_\kk^\pm \propto
e^{i\omega \eta}$, we recover the well-known result that two
polarization states of photons propagate with different phase
velocities, causing the polarization plane to rotate by an angle
$\Delta \phi = -(1/2) \! \int \! d\eta \, m(\eta)$ (to the first
order in $m/k$)~\cite{Carroll}.

The study of this effect on cosmological scales, known as {\it
cosmological birefringence}, gives the most stringent bounds on
$m$~\cite{Carroll,Kostelecky-q}. In fact, $\Delta \phi$ has been
constrained by observation of radio galaxies and quasars, $\Delta
\phi(z_{\rm 3C9}) = 2^{\circ} \pm 3^{\circ}$ ($68\%$
C.L.)~\cite{Carroll2} (where 3C9 at $z_{\rm 3C9} = 2.012$ is the
best data sets available at a single redshift), and by the
analysis of Cosmic Microwave Background (CMB) polarization
fluctuations~\cite{WMAP}. In the latter case, $\Delta \phi$ has
been constrained between the reionization epoch ($z_{\rm reion}
\sim 10$) and today, $-22.2^{\circ} < \Delta \phi(z_{\rm reion}) <
7.2^{\circ}$ ($95\%$ C.L.), and between the decoupling epoch
($z_{\rm dec} \simeq 1090$) and today, $-5.5^{\circ} < \Delta
\phi(z_{\rm dec}) < 3.1^{\circ}$ ($95\%$ C.L.). The above limits
give, with $m$ constant in time, the bound $m \lesssim 10^{-34}
\eV$.
However, this bound refers to relatively recent epochs and it is
not excluded that in the early universe the value of $m$ could
have been much greater. In order to avoid inessential
complications, we assume that Lorentz symmetry is broken at some
energy scale before or during inflation, and that $m$ is constant
during inflation and decreases afterward to fulfil the above
constraint.

The solution of Eq.~(\ref{Autoinduction}) with $m$ constant is
%
%
\begin{equation}
\label{AutoinductionSol} a^2 B_\kk^\pm = \pm k \exp[ik\eta \sqrt{1
\pm m/k} \:]/\sqrt{2k} \, .
\end{equation}
Here, the normalization is such that for $\eta \rightarrow
-\infty$ (where $m$ is supposed to be zero) we have the usual
Maxwell vacuum spectrum, $a^2 B_\kk^\pm = \pm k
e^{ik\eta}/\sqrt{2k}$. 
From Eq.~(\ref{AutoinductionSol}) we easily deduce that only
negative-helicity modes such that $1/(m \eta^2) \ll k \lesssim m$
are (exponentially) amplified and that in this case $a^2 B_\kk^+
\simeq k \exp(k|\eta|\sqrt{m/k})/\sqrt{2k}$. Using
Eq.~(\ref{Blambda1-H2}), 
the amplitude of the (almost) maximally helical field 
at the end of inflation, $\eta = \eta_{\rm rh}$, reads 
%
%
\begin{equation}
\label{BlambdaRH-HlambdaRH} a_{\rm rh}^2 B_\lambda(\eta_{\rm rh})
\simeq e^\gamma/2\pi \lambda^2 \! \sqrt{\gamma} \, ,
\end{equation}
where $a_{\rm rh} = a(\eta_{\rm rh})$, $\gamma = (m \eta_{\rm
rh}^2/\lambda)^{1/2}$, and we assumed $1/m \lesssim \lambda \ll m
\eta_{\rm rh}^2$~\cite{note3}. Looking at
Eq.~(\ref{BlambdaRH-HlambdaRH}), we see that the magnetic field is
peaked at small scales since the exponential factor is inversely
proportional to $\sqrt{\lambda}$. Therefore, at the end of
inflation, magnetic power is highly concentrated in scales just a
little greater than $\lambda \simeq 1/m$ and negligible at larger
scales. In this case, the condition $\lambda \ll m \eta_{\rm
rh}^2$ (or equivalently $\gamma \gg 1$) reads $m |\eta_{\rm rh}|
\gg 1$. This means that the mode corresponding to $\lambda \simeq
1/m$ is well inside the horizon since its physical wavelength is
much smaller than the Hubble radius $H^{-1}$.

After inflation, the universe enters in the so-called reheating
phase, during which the energy of the inflaton is converted into
ordinary matter. In this paper, we restrict our analysis to the
case of instant reheating, that is we assume that after inflation
the universe enters the radiation dominated era. In this era, as
well as in the subsequent matter era, the effects of the
conducting primordial plasma are important when studying the
evolution of a magnetic field. They are taken into account by
adding to the Maxwell action the source term $\int \! d^4 x \, e
j^{\mu} \! A_{\mu}$, where $j^{\mu} = (0, \sigma_c {\textbf E})$
is the external current~\cite{Turner} (we are assuming that the
Chern-Simons term is negligible after inflation). Plasma effects
introduce, in the left-hand-side of Eq.~(\ref{Autoinduction}), the
term $-a\sigma_c \dd_\eta (a^2 B_\kk^\pm)$.

For modes well inside the horizon, $k|\eta| \gg 1$,
Eq.~(\ref{Autoinduction}) reduces to $k^2 a^2 B_\kk^\pm \simeq
-a\sigma_c \dd_\eta (a^2 B_\kk^\pm)$ so that $a^2 B_\kk^\pm(\eta)
= a_{\rm rh}^2 B_\kk^\pm(\eta_{\rm rh}) \exp[- \! \int \! d\eta
(k^2/a\sigma_c)]$. Introducing the comoving {\it dissipation
length}, $\xi_{\rm diss}^2 = \int \! d \eta (1/a\sigma_c) \simeq
1/(a^2 H\sigma_c)$, where we used the relation $\eta \simeq
1/(aH)$ valid in all eras, we get $a^2 B_\kk^\pm(\eta) = a_{\rm
rh}^2 B_\kk^\pm(\eta_{\rm rh}) \exp(-\xi_{\rm diss}^2 k^2)$. This
means that modes with comoving wavelength greater than the
comoving dissipation length are frozen into the plasma, while
modes for which $\lambda \lesssim \xi_{\rm diss}$ are washed out.
In particular, only modes with comoving wavelength greater than
the actual dissipation length, $\xi_{\rm diss}(T_0) \simeq 1
\AU$~\cite{Widrow}, can survive until today
($T_0 \simeq 2 \times 10^{-4} \eV$ 
is the actual temperature~\cite{Kolb}). But this is not our case
since wavelengths greater than $1 \AU$ would correspond to modes
outside the horizon at the end of inflation. Thus, we find the
uninteresting result that the inflation-produced field is
dissipated soon after the beginning of radiation era.

This situation drastically changes if one takes into account that
the primordial plasma could have been turbulent~\cite{Widrow}.
Indeed, turbulence naturally arises during (first-order)
cosmological phase transitions and those could have affected the
universe during its evolution. In the following we simply assume
that the early universe is turbulent without referring to any
specific, generating model of turbulence (if this is not the case,
our mechanism cannot account for the presence of large-scale
magnetic fields in galaxies, as discussed above).

At the beginning of radiation era, $T = T_{\rm rh}$, and at the
comoving scale $1/m$ the kinetic and magnetic Reynolds numbers are
$\mbox{Re} \simeq \alpha v T_{\rm rh}/m$ and $\mbox{Re}_B \simeq v
T_{\rm rh}/(\alpha m)$, where we used $\nu \simeq 1/(\alpha T)$,
$\sigma_c \simeq T/\alpha$~\cite{Son}, and $\alpha$ is fine
structure constant. For an incompressible, relativistic turbulent
plasma an upper limit to $v$ is the sound speed $v_{\rm s} =
1/\sqrt{3}$. Taking, for example, $T_{\rm rh} \simeq 10^8 \GeV$,
$m \simeq 10^{-13} \eV$, and $v \simeq v_{\rm s}$, we get
$\mbox{Re} \simeq 10^{27}$ and $\mbox{Re}_B \simeq 10^{32}$. The
temperature at the beginning of radiation era, the so-called {\it
reheat temperature}, is constrained in the range $1\GeV \lesssim
T_{\rm rh} \lesssim 10^8 \GeV$
from analyses of Big Bang Nucleosynthesis (BBN) and CMB
radiation~\cite{Riotto}. For instant reheating, it results $T_{\rm
rh} \simeq (H \mPl)^{1/2}$~\cite{Turner}, where $H$ is the Hubble
parameter during inflation and $\mPl \simeq 10^{19} \GeV$ 
the Planck mass.

In a spatially-flat Friedman universe, the decay of a turbulent,
maximally helical field proceeds through inverse cascade and the
evolution laws are equal to those in a static Minkowski universe
provided that time, intensity and physical correlation length of
the magnetic field are replaced by the conformal time, the
``comoving intensity'' $a^2 B_\xi$ and comoving length,
respectively~\cite{Son}. Therefore, we have $a^2 B_\xi \propto
\xi^{-1/2}$ and $\xi(\eta) \simeq \xi(\eta_{\rm rh})
[\eta/\tau_{\rm eddy}(\eta_{\rm rh})]^{2/3}$, where $\tau_{\rm
eddy}(\eta_{\rm rh}) = \xi(\eta_{\rm rh})/v(\eta_{\rm rh})$.
The inverse cascade terminates when the primordial plasma ceases
to be turbulent. This happens when the universe has sufficiently
cooled down to render kinematic viscosity very large, thereby
causing the kinetic Reynolds number to drop below unity at the
scale of interest. From this point on, the evolution of the
magnetic field is adiabatic, in the sense that $a^2 B_\xi =
\mbox{const}$ and $\xi = \mbox{const}$. Taking into account that
for $T \ll 1 \MeV$ the kinematic viscosity is
$\nu \simeq 10^{11} \MeV^2 \! / \,T^3$~\cite{Son} 
and that $v$ scales in time as the comoving magnetic field,
$v(\eta) \simeq v(\eta_{\rm rh}) [\tau_{\rm eddy}(\eta_{\rm
rh})/\eta]^{1/3}$~\cite{Son},
we obtain the temperature at which the plasma becomes
non-turbulent [starting at the time $\eta_{\rm rh}$ with
$v(\eta_{\rm rh}) \simeq v_{\rm s}$ and $\xi(\eta_{\rm rh}) \simeq
1/m$]: $T_{\rm nt} \simeq 32 m_{13}^{1/4} \eV$
or $T_{\rm nt} \simeq 28 m_{13}^{4/17} \eV$
according to $m_{13} \gtrsim 6 \times 10^{-5}$ or $m_{13} \gtrsim
6 \times 10^{-5}$, where $m_{13} = m/10^{-13} \eV$. The two cases
correspond, respectively, to have $T_{\rm nt} \gtrsim T_{\rm eq}$
or $T_{\rm nt} \lesssim T_{\rm eq}$, where
$T_{\rm eq} \simeq 3 \, \eV$ 
is the temperature at the matter-radiation equality~\cite{Kolb} .

Evolving (along the lines discussed above) the magnetic
field~(\ref{BlambdaRH-HlambdaRH}) evaluated at the scale $1/m$,
from the end of inflation till the time of protogalactic collapse,
we get
\begin{eqnarray}
\label{B0} && \!\!\!\!\!\!\!\!\!\!\!\! B_\xi(t_{\rm pg}) \simeq 3 
\times 10^{-31} \! \left[\frac{10\kpc}{\xi_{\rm phys}(t_{\rm
pg})}\right]^{\!1/2} \! m_{13} T_8^{1/2} e^\gamma \, \G,
\\
\label{xi0} && \!\!\!\!\!\!\!\!\!\!\!\! \xi_{\rm phys}(t_{\rm pg})
\simeq
    \left\{
            \begin{array}{ll}
            0.1 \, m_{13}^{-1/2} \, \pc   & \mbox{if} \;\, m_{13} \gtrsim 6 \times 10^{-5},
            \\
            0.3 \, m_{13}^{-7/17} \, \pc  & \mbox{if} \;\, m_{13} \lesssim 6 \times 10^{-5},
            \end{array}
    \right.
\end{eqnarray}
where $\gamma \simeq m |\eta_{\rm rh}| \simeq 52 \, m_{13}
T_8^{-1}$ 
and $T_8 = T_{\rm rh}/10^8 \GeV$. In order to explain the
existence of galactic magnetic fields, we impose $B_\xi(t_{\rm
pg}) \simeq 10^{-6} [10 \kpc/\xi_{\rm phys}(t_{\rm pg})]^{1/2}
\G$. From Eq.~(\ref{B0}), it is easy to see that the above
requirement is satisfied for $m_{13} \simeq T_8$, 
so we find that the magnetic field at the time of protogalactic
collapse has a physical correlation length ranging from $0.1 \pc$
to $0.6 \kpc$ and then could be exactly the field detected in
galaxies~\cite{note5}.

Finally, if we assume as in Ref.~\cite{Bertolami} that $m$ scales
as a power of temperature, $m(T) \propto T^l$, and take $l=1$
(i.e. $m$ simply redshifts after inflation), we find that its
present-day value is $m_0 \simeq
2 \times 10^{-34} \eV$, 
and that the rotation angle of polarized light emitted from a
source at a redshift $z_i$ and detected today is $\Delta \phi(z_i)
\simeq (3/2)t_0 m_0 [1-\sqrt{1+z_i}\,]$. For the three cases
discussed above, we get $\Delta \phi(z_{\rm 3C9}) \simeq
-8^{\circ}$, $\Delta \phi(z_{\rm reion}) \simeq -26^{\circ}$, and
$\Delta \phi(z_{\rm dec}) \simeq -362^{\circ}$ (equivalent to
$-2^{\circ}$) which shows, roughly speaking, that $m$ is
consistent with observational constraints for $l \gtrsim 1$.


It is worth stressing that, although other mechanisms for
generating cosmic magnetic fields exist in the
literature~\cite{Widrow}, 
the field produced in our mechanism has three features which are
not simultaneously present in any of them: it directly explains
galactic magnetism (without resorting to further and still-debated
amplification mechanisms such as {\it galactic
dynamos}~\cite{Widrow,note6}, it is maximally helical, and its
power is concentrated on a single scale, the magnetic correlation
length. These characteristics make it peculiar and hopefully
detectable in future observations of CMB
temperature and polarization fluctuations~\cite{Giovannini}. 


In conclusion, we have studied the generation of a cosmic magnetic
field in Maxwell-Chern-Simons theory of Electrodynamics containing
a Lorentz-violating coupling between the photon and an external
four-vector. The breaking of conformal invariance in this theory
allow electromagnetic vacuum fluctuations to be amplified during
inflation. The resulting field at the end of (de Sitter) inflation
is maximally helical and is subsequently processed by
magnetohydrodynamic effects taking place in the primeval turbulent
plasma. In particular, quasi-conservation of magnetic helicity is
responsible for an inverse cascade of magnetic spectrum which
causes a growth of the correlation length together with a slow
dissipation of the magnetic field intensity. As a result, for
appropriate magnitude of the four-vector, the produced field at
the time of protogalactic collapse has an intensity and
correlation length of the right magnitude in order to explain
galactic magnetism.


\begin{acknowledgments}
We thank V.~A.~Kosteleck\'{y} for valuable comments.
\end{acknowledgments}




\begin{thebibliography}{99}
\bibitem{note1}          Due to the very high temperature and very low density
                         of the collapsing cloud, the protogalactic gas is fully ionized.
                         The mass density $\rho_{\rm pg}$ refers to this component
                         of the protogalaxy and does not take into account
                         the ``dark matter'' component which constitutes about the
                         $90\%$ of the protogalactic mass. 
                         In this paper, we assume that the total mass of a
                         typical protogalaxy is $10^{11}$ solar masses.
\bibitem{R1}             It is worth noting that, after a galaxy forms owing to protogalaxy collapse,
                         the typical velocity of turbulent plasma is no longer given by
                         the virial thermal speed but is determined primarily by
                         astrophysical mechanism such as supernovae explosions
                         and differential rotation of the galactic disk.
\bibitem{R2}             Estimates based on ``spherical infall model'' indicate that
                         $z_{\rm pg} \simeq 50$. However, we
                         stress here that galactic disks form at a
                         much later ephoch, at a redshift of order
                         $z \sim$ few.
\bibitem{note3}          In order to have simple formulae, we approximated
                         $\int_0^{\infty} \!\! dk k^3 e^{-k^2 \lambda^2} e^{2|\eta_{\rm rh}| \sqrt{mk}}$
                         with $\lambda^{\!-3} \! \int_0^{1/\lambda} \! dk \, e^{2|\eta_{\rm rh}| \sqrt{mk}}$.
\end{thebibliography}
\end{document}